\documentclass[aps,prl,twocolumn,superscriptaddress,nofootinbib]{revtex4-1}

\usepackage{mathtools}
\usepackage{lipsum}
\usepackage{amsmath,amssymb}
\usepackage{MnSymbol}
\usepackage{graphicx}
\usepackage{color}
\usepackage{hyperref}
\usepackage{url}
\usepackage{slashed}
\usepackage{slashed,bbm}
\usepackage{graphics,psfrag,epsfig}
\usepackage{dsfont}
\usepackage{enumerate}
\usepackage{pbox}
\usepackage{braket}
\usepackage[retainorgcmds]{IEEEtrantools}
\usepackage{xr}
\usepackage{gensymb}
\usepackage{multibib}

\DeclareMathSizes{10}{10}{6}{4}

\begin{document}
	\title{Time-Reversal Invariant Topological Superconductivity \\in Planar Josephson Bijunction}
	\author{Yanick Volpez}
	\affiliation{Department of Physics, University of Basel, Klingelbergstrasse 82, CH-4056 Basel, Switzerland}	
	\author{Daniel Loss}
	\affiliation{Department of Physics, University of Basel, Klingelbergstrasse 82, CH-4056 Basel, Switzerland}
	\author{Jelena Klinovaja}
	\affiliation{Department of Physics, University of Basel, Klingelbergstrasse 82, CH-4056 Basel, Switzerland}
\date{\today}	
	
\begin{abstract}
We consider a Josephson bijunction consisting of a thin $SIS$ $\pi$-Josephson junction sandwiched between two-dimensional semiconducting layers with strong Rashba spin-orbit interaction. Each of these layers forms an $SNS$ junction due to proximity-induced superconductivity. The $SIS$ junction is assumed to be thin enough such that the two Rashba layers are tunnel-coupled. We show that, by tuning external gates, this system can be controllably brought into a time-reversal invariant topological superconducting phase with a Kramers pair of Majorana bound states being localized at the end of the normal region for a large parameter phase space. In particular, in the strong spin-orbit interaction limit, the topological phase can be accessed already in the regime of small tunneling amplitudes.
\end{abstract}
	\maketitle

{\it Introduction.} The prediction of Majorana bound states (MBSs) in $p$-wave superconductors \cite{Kitaev2001} and the proposed implementation in semiconductor-superconductor heterostructures \cite{Lutchyn2010,Oreg2010,Alicea2010} has sparked wide interest in exploring new platforms hosting topological superconductivity. Prominent experimental realizations involve nanowires proximitized by an $s$-wave superconductor \cite{Chang2015,Kjaergaard2016,Shabani2016,Gazibegovic2017,Lutchyn2018,Vaitiekenas2018}, or chains of magnetic adatoms on $s$-wave superconductors \cite{mf1,mf2,mf3,mf4,mf5,mf6,mf7,mf8,mf9}. More recently, MBSs in Josephson junctions \cite{Pientka2017} gained considerable interest \cite{Hell2017,Hell20172,Hart2017,Nichele2017,Yacoby2018,Zhi2019,Lee2019,Setiawan2019,Mitra2018,Nichele2018,Scharf2019}, since this setup offers additional control knobs for experiments, in particular, the superconducting phase difference. In this case, the topological phase can be reached for a wide parameter range and the scalability to topological networks seems promising \cite{Pientka2017}. 

In contrast, time-reversal invariant (TRI) topological superconductors hosting Kramers pairs of MBSs \cite{Fu2008,Sato2009,Qi2009,Fu2010,Trauzettel2011,meng,schulz,Nakosai2012,Dmytruk2019,kl2015,Deng2012,Zhang2013,Parhizgar2016,Wang2016,JK2014,Arra1,Flensberg2014,Schrade2017,Wang2018,Hsu2018,Yan2018,Schrade2018,Haim2014,Camjayi2017} are still lacking experimental evidence. Such schemes are attractive since they avoid magnetic fields and their detrimental effect on superconductivity. Many such proposals rely on unconventional superconductors or can only be achieved with strong electron-electron interactions \cite{Haim2016,Manisha2018}. A major drawback of them is that they do not allow for an easy {\it in-situ} control over the topological phase. 

	\begin{figure}[t]
		\centering
		\includegraphics[width=0.8\columnwidth]{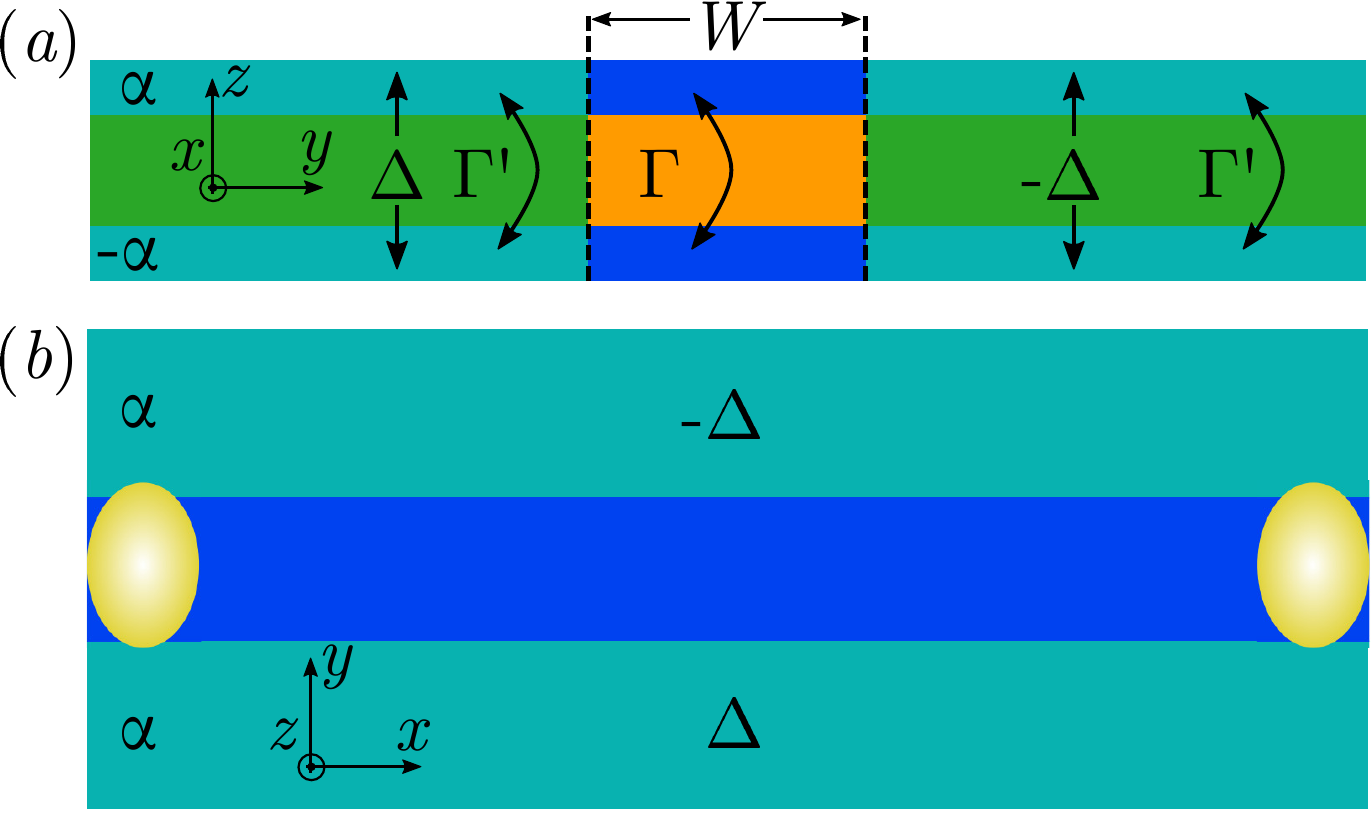}
		\caption{(a) 
		Josephson bijunction setup consisting of 
		a top (blue and mint) and bottom (blue and mint) Rashba layer with opposite SOI in contact with an $SIS$ junction (green-orange-green). The superconductors (green) have a phase difference $\pi$ and are separated by an insulator (orange) of width $W$. The proximity to the $SIS$ junction creates two $SNS$ junctions of width $W$ in the Rashba layers. (b) Top view of the setup. In the topological phase a Kramers pair of MBSs (yellow dots) emerges at each end of the junction.}
		\label{Setup}
	\end{figure}

It is the goal of the present work to close this gap. For this we propose a Josephson bijunction
which can be brought into the TRI topological superconducting phase via the control of experimentally accessible parameters, thereby hosting multiple Kramers pairs of MBSs, see Fig. \ref{Setup}. The setup consists of a superconductor-insulator-superconductor ($SIS$) Josephson junction sandwiched between semiconducting layers with Rashba spin-orbit interaction (SOI) [see Fig. \ref{Setup}(a)]. The superconductors are assumed to be $s$-wave and to have a phase difference $\phi$ controlled by applying a magnetic flux or generated by introducing an additional layer of randomly oriented magnetic impurities \cite{Buzdin,Ryazanov,Spivak,Dam, Schrade2015}. Due to the proximity effect this leads to the formation of two tunnel-coupled superconductor-normal-superconductor ($SNS$) Josephson junctions of width $W$ and phase difference $\phi$ [see Fig. \ref{Setup}(a)]. This setup leads to an intricate interplay between the formation of Andreev bound state bands (ABSBs) in both $SNS$ junctions on one hand and the hybridization of the two Rashba layers on the other hand, with striking consequences. In particular, we find a periodic closing and reopening of the topological gap in the spectrum of the ABSBs as a function of $k_{so}W$, where $k_{so}$ is the SOI momentum which can be tuned electrically via gates 
\cite{so1,so2,so3}. Interestingly, in the strong SOI limit, the TRI topological phase can be reached for a relatively small tunneling amplitude.

We believe that with the recent advances in the fabrication and study of bilayer systems \cite{Pablo2018_1,Pablo2018_2,Yankowitz2019,Yazdani2019} and van der Waals heterostructures \cite{Novoselov2016,Novoselov2013,Novoselov2012,Liu2016,Geim2013}, the setup proposed here lies within experimental reach and could be a promising route in realizing Kramers pairs of MBSs without the need of magnetic fields.

{\it Model.} The Josephson bijunction consists of two semiconducting layers 
labeled by $\tau=\pm1$. 
Due to the structural asymmetry the Rashba SOI vectors ${\boldsymbol \alpha}_{\tau}$ are naturally antiparallel and aligned along the $z$ axis, which is normal to the layers. The layers are modeled by 
\begin{equation}
H_{\tau} = \sum_{\sigma, \sigma'} \int d\boldsymbol{r} \ \psi^{\dagger}_{\tau \sigma}(\boldsymbol{r}) \mathcal{H}_{\tau \sigma \sigma'}(\boldsymbol{r}) \psi_{\tau \sigma'}(\boldsymbol{r}),
\end{equation}
where $\mathcal{H}_{\tau \sigma \sigma'}(\boldsymbol{r})=[-\hbar^2 \nabla^2/(2m)- \mu_{\tau}+ \boldsymbol{\alpha}_{\tau} \cdot \boldsymbol{\sigma} \times \hat{\mathbf{k}}]_{\sigma \sigma'}$, $m$ is the effective mass, $\mu_{\tau}$ the chemical potential, $\hat{\mathbf{k}}=-i \hbar (\partial_x,\partial_y)$ the in-plane momentum operator, and $\sigma_i$ are the Pauli matrices acting in spin space. The operator $\psi^{\dagger}_{\tau \sigma}(\boldsymbol{r})$ creates an electron with spin projection $\sigma$ along the $z$ axis in the $\tau$-layer at the position $\boldsymbol{r}=(x,y)$. The SOI energy $E_{\tau,so}= \hbar^2 k_{so,\tau}^2/2m$, with the SOI momentum $k_{so,\tau}=m \alpha_{\tau}/\hbar^2$, is the energy difference between the bottom of the Rashba bands and the degeneracy point at $k=0$, from which $\mu_{\tau}$ is measured.

\begin{figure}
	\centering
	\includegraphics[width=\columnwidth]{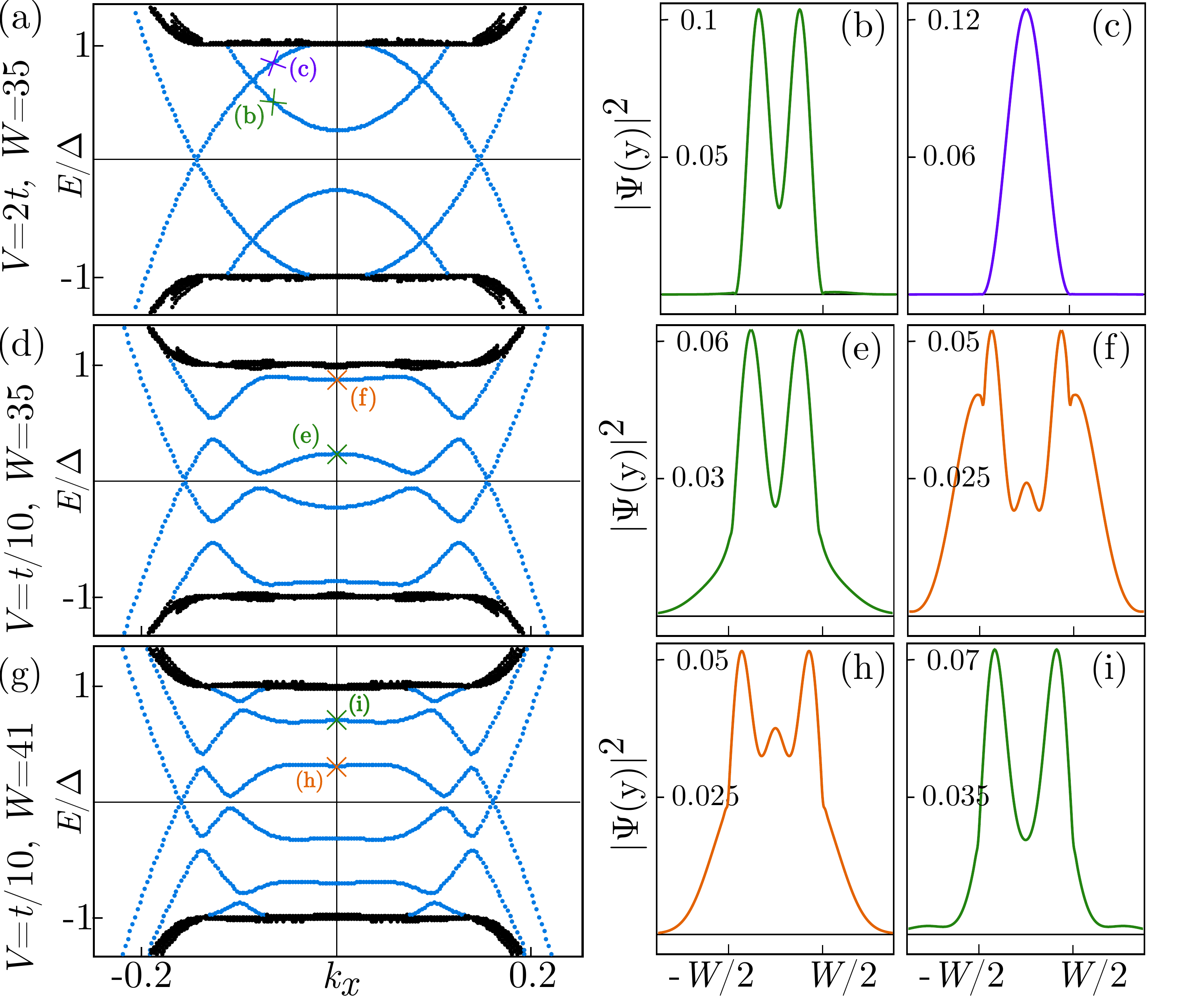}
	\caption{Spectrum as function of momentum $k_x$ for (a) $V=2t$ and $W=35$, (d) $V=t/10$ and $W=35$, and (g) $V=t/10$ and $W=41$. (a) For a high potential barrier $V$, the transverse subbands [$n=1$ (purple cross) and $n=2$ (green cross)] are decoupled from the superconducting regions and lie within the superconducting gap (black). (b)-(c) The corresponding probability densities $|\psi(y)|^2$ at $k_x=-0.08$ of the first two transverse subbands. (d) Upon lowering $V$, the interfaces become transparent leading to ABSBs. The electron and hole bands are mixed now and as a consequence each ABSB has contributions from different transverse subbands with the dominant orbital content depending on $k_x$. The lowest ABSB also has a very small gap around $|k_x| \approx 0.2$, which is further discussed in the text. (e)-(f) At $k_x=0$, the lowest (second) ABSB comes from the $n=2$ ($n=3$) subband of the normal region. (g) Between $W=35$ and $W=41$, the inverted case emerges where the lowest (second) ABSB comes from the $n=3$ ($n=2$) subband [(h) and (i)]. This is the origin of the topological phase transition of higher subbands (see text below and SM \cite{SM}). Numerical parameters are $a=1$, $N_y=500$, $N_W=W-1$, $\alpha_1/t=-\alpha_{\bar{1}}/t=0.07$, $\Delta/t=0.012$, $\Gamma=1.5\Delta$, and $\Gamma'=0$.}
	\label{Spectra_transverse}
\end{figure}

The regions that are in contact with the bulk superconductors become superconducting as well \cite{Pientka2017,Nichele2018,Yacoby2018,Scharf2019} and the proximitized Rashba layers themselves form an $SNS$ junction with a normal region of width $W$ \cite{reeg1}. The superconducting regions are described by \cite{Footnote1}
\begin{equation}
H_{\Delta} = \sum_{{\tau, \sigma, \sigma'}} \int d\boldsymbol{r} \ \Big[ \Delta_{\tau}(y) \psi^{\dagger}_{\tau \sigma}(\boldsymbol{r}) [i \sigma_2]_{\sigma \sigma'} \psi^{\dagger}_{\tau \sigma'}(\boldsymbol{r}) + \text{H.c.}\Big], \label{HamiltonianSC}
\end{equation}
where $\Delta_{\tau}(y)=\Delta e^{i \text{sgn}(y)\phi/2}\theta(|y|-W/2)/2$, $\Delta>0$ is the strength of the induced superconducting pairing and $\phi$ is the phase difference between the superconducting regions inherited from the parent superconductors. We also allow for electron tunneling between the two layers,
\begin{equation}
H_{\Gamma} = \sum_{\sigma} \int d\boldsymbol{r} \ \Big[ \bar{\Gamma}(y) \psi^{\dagger}_{1 \sigma}(\boldsymbol{r}) \psi_{\bar{1} \sigma}(\boldsymbol{r}) + \text{H.c.} \Big],
\end{equation}
where we assume the tunneling amplitude to be uniform in the $x$ direction such that $\bar{\Gamma}(y)=\Gamma \theta(W/2-|y|) + \Gamma' \theta(|y|-W/2)$, and without loss of generality $\Gamma>0$ ($\Gamma' \geq 0$) for the coupling between the two layers \cite{reeg2}. The total Hamiltonian modeling the setup is then given by $H=H_{1}+H_{\bar{1}}+H_{\Delta}+H_{\Gamma}$. 

First, we study the system numerically for two geometries [see the Supplemental Material (SM) \cite{SM} for details]: ($i$) {\it semi-infinite geometry} - the system is translationally invariant along the direction of the junction, {\it i.e.}, along the $x$-direction, which allows us to parametrize the eigenstates by the good quantum number $k_x$; ($ii$) {\it finite geometry} - the system is finite in both the $x$- and $y$-direction. 
Second, to gain a better physical understanding, we then treat the problem also analytically.

{\it Spectrum of Andreev Bound State Bands.} In order to better understand the evolution of the ABSBs, it is instructive to consider the effect of a potential barrier of height $V$ at the interfaces between the superconducting and the normal regions ($y=\pm W/2$ ) \cite{Olesia2018}. If $V$ is very large, the interfaces are intransparent such that there are two independent superconducting regions and an isolated normal region, which consists of two tunnel-coupled quantum wires. Since the physics of the normal region in the confined direction ($y$-direction) is equivalent to a particle in a box, the spectrum consists of quantized transverse subbands labeled by an index $n$, where the spacing between them depends on the width $W$ and roughly behaves as $ 1/W^2$. Out of these subbands only the ones crossing the chemical potential or lying within the superconducting gap are important for us. By increasing $W$ the subband spacing is reduced and as a result more of these will be shifted into the superconducting gap [see Fig. \ref{Spectra_transverse}a]. The probability density of the $n$-th subband, $|\psi^{(n)}(y)|^2$, has $n$ peaks, which allows us to determine $n$ numerically from the shape of the wavefunction.

Upon lowering $V$, the electron and hole bands get coupled with corresponding anti-crossings and, as a result, ABSBs form [see Fig. \ref{Spectra_transverse}b]. This picture qualitatively explains why the number of ABSBs increases with $W$, and that generically the ABSBs have contributions coming from different transverse subbands of the normal region [see Fig. \ref{Spectra_transverse}]. For $\Gamma \neq 0$, the ABSB spreads in both layers. More concretely, when a new subband is entering the superconducting gap, the ABSB highest in energy will obtain mainly contributions from this additional subband around $k_x=0$, while at larger momenta $|k_x|$ it has mainly contributions from a subband with smaller $n$. As $W$ is further increased the same behavior can be observed for all ABSBs [see Fig. \ref{Spectra_transverse}]. For the remainder we will put $V=0$ and work with ABSBs. 

\begin{figure}
	\centering
	\includegraphics[width=\columnwidth]{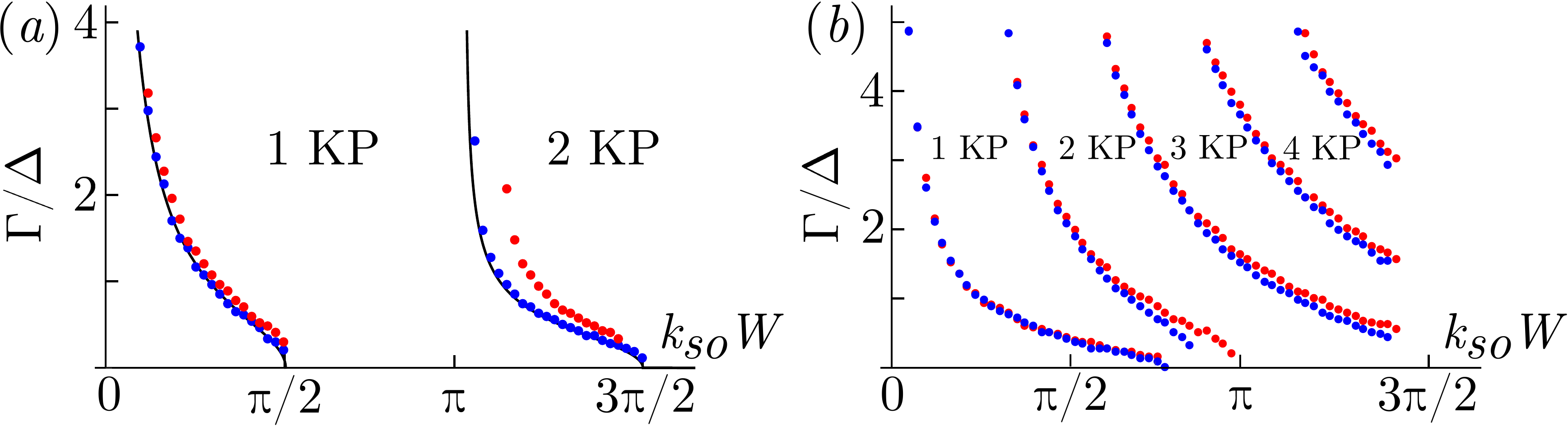}
	\caption{Phase diagram as function of $k_{so}W$ and tunneling amplitude $\Gamma$ for (a) strong SOI ($\Delta=E_{so}/20$) and (b) weak SOI ($\Delta \approx E_{so}$). 
		After each gap closing at $k_x=0$, a new Kramers pair of MBSs emerges. For strong SOI, we find very good agreement between numerical and analytical [black line, see Eq. \eqref{GapClosingCondition}] results. The numerical results (blue dots) are obtained in the semi-infinite geometry and parameters are $N_y=10000$, $\alpha_1/t=-\alpha_{\bar{1}}/t=0.07$. In the generic case $|\alpha_1| \neq |\alpha_{\bar{1}}|$, the gap closing line still shows a similar behavior (red dots). Here, we choose $\alpha_1/t=0.07$ and $\alpha_{\bar{1}}/t=-0.06$.}
	\label{PhaseDiag}
\end{figure}

{\it Kramers pairs of MBSs.} If the phase difference between the superconductors is $\phi=\pi$, the system is time-reversal invariant and is in the DIII symmetry class and has a $\mathbb{Z}_2$ number classification \cite{Ryu2010}. The time-reversal (particle-hole) symmetry operator is given by $\Theta= i \sigma_2 \mathcal{K}$ ($\mathcal{P}=\eta_1 \mathcal{K}$), where $\eta_i$ are the Pauli matrices acting in particle-hole space, and $\mathcal{K}$ is the complex conjugation operator.

To investigate topological phase transitions, we look for gap closings of the ABSBs at $k_x=0$, which only happens at particular combinations of $k_{so}W$ and $\Gamma/\Delta$. An analytical expression for the gap closing condition can be derived in the semi-infinite geometry in the strong SOI limit, {\it i.e.} $\Gamma, \Delta \ll E_{so}$, for $\alpha_1=-\alpha_{\bar{1}}$ and $\Gamma'=0$ (see SM \cite{SM} for details). In this limit, we find a remarkably simple expression for the gap closing condition
\begin{equation}\label{GapClosingCondition}
\left(\frac{\Gamma}{\Delta}\right)^2- \frac{\cot(k_{so}W)}{k_{so}W}=0.
\end{equation}
We stress that it is essential for obtaining this result to go beyond standard linearization and retain band curvature effects. Thus, for a fixed ratio $\Gamma/\Delta$ there exist multiple solutions, whenever the cotangent is positive [see Fig. \ref{PhaseDiag}a]. In the interval where the cotangent is negative, {\it e.g.} for $k_{so}W \in (\pi/2,\pi)$, no gap closing occurs and the ABSBs are always gapped [see Fig. \ref{PhaseDiag}a]. Interestingly, in this region of the phase diagram the topological phase can in principal be achieved with an infinitesimal tunneling amplitude $\Gamma$. Relaxing the condition of exactly opposite SOI in the two layers, one finds that the functional form of the gap closing lines remain essentially unchanged [see Fig. \ref{PhaseDiag}]. However, for weak SOI, $\Delta \sim E_{so}$, the functional form deviates from the cotangent behavior, and, more importantly, there are no longer regions where no gap closing takes place [see Fig. \ref{PhaseDiag}b]. Note that the position of the gap closing lines depend on the product $k_{so}W$, and it is thus possible to tune into a particular phase by changing the SOI $\alpha$, {\it e.g.} by applying electric fields via external gates.
\begin{figure}
	\centering
	\includegraphics[width=\columnwidth]{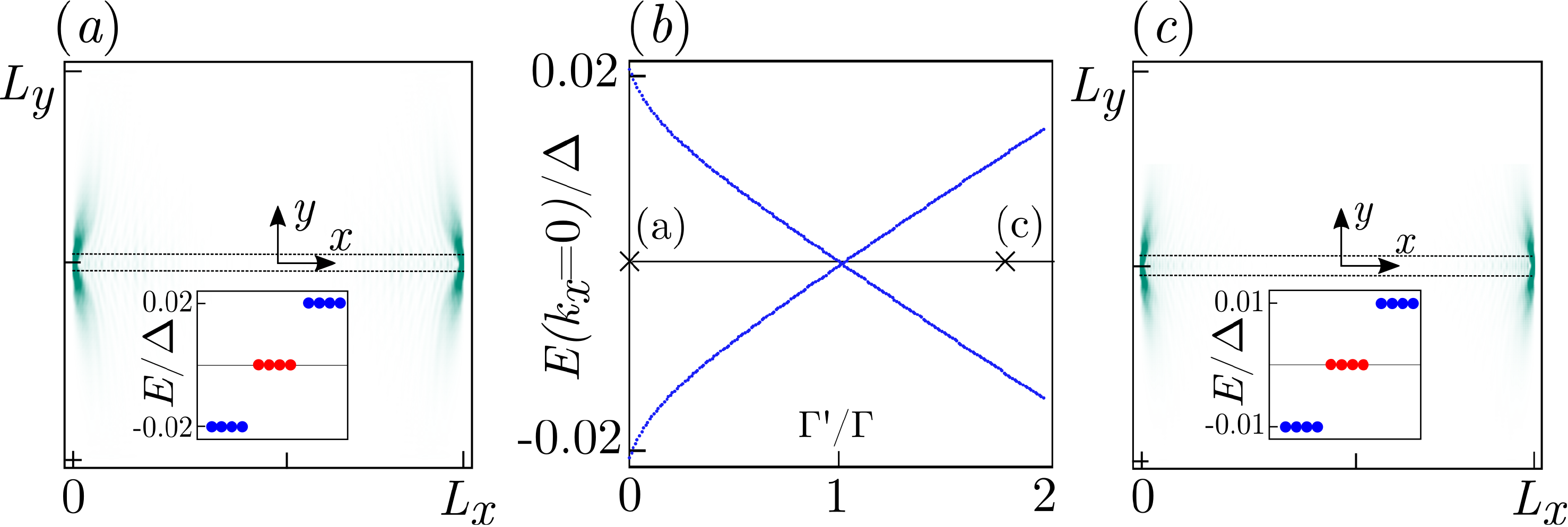}
	\caption{Color plot of probability density $|\psi(x,y)|^2$ for two MBSs localized at opposite ends of the junctions for (a) $\Gamma'=0$ and (c) $\Gamma'/\Gamma=1.8$. In total, there are four zero-energy MBSs (red dots in spectrum in inset), with two Kramers pairs of MBSs at each end of the junction. The dashed lines symbolize the extension of the normal region. (b) Spectrum at $k_x=0$ in semi-infinite geometry as function of $\Gamma'/\Gamma$. The gap in the ABSB closes around $\Gamma' \approx \Gamma$ and reopens. The crosses mark the values of $\Gamma'/\Gamma$ chosen for panels (a) and (c). A Kramers pair of MBSs exists for both values and the topological phase can thus be reached for both regimes $\Gamma>\Gamma'$ and $\Gamma<\Gamma'$. Parameters are $N_x=800$, $N_y=400$, $N_W=6$, $\alpha_1/t=-\alpha_{\bar{1}}/t=0.35$, $\Delta=E_{so}/10$, and $\Gamma=0.6\Delta$.}
	\label{MBS_TRS}
\end{figure}

The closing and reopening of the gap of the ABSBs is accompanied by the emergence of a Kramers pairs of MBSs at each end of the normal region [see Fig. \ref{MBS_TRS}], thus these gap closing points mark topological phase transitions. Due to their topological nature they are robust against potential disorder, as we checked numerically. Importantly, the emergence of the MBSs does neither rely on the fine-tuned point $|\alpha_1|=|\alpha_{\bar{1}}|$ nor on $\Gamma'=0$. It turns out that MBSs exist for both regimes $\Gamma >\Gamma'$ and $\Gamma<\Gamma'$ [see Fig. \ref{MBS_TRS}], which implies that the relative strength $\Gamma/\Gamma'$ is not crucial to reach the topological phase. However, it is essential that $\Gamma$ is non-zero.
\begin{figure}
	\centering
	\includegraphics[width=.85\columnwidth]{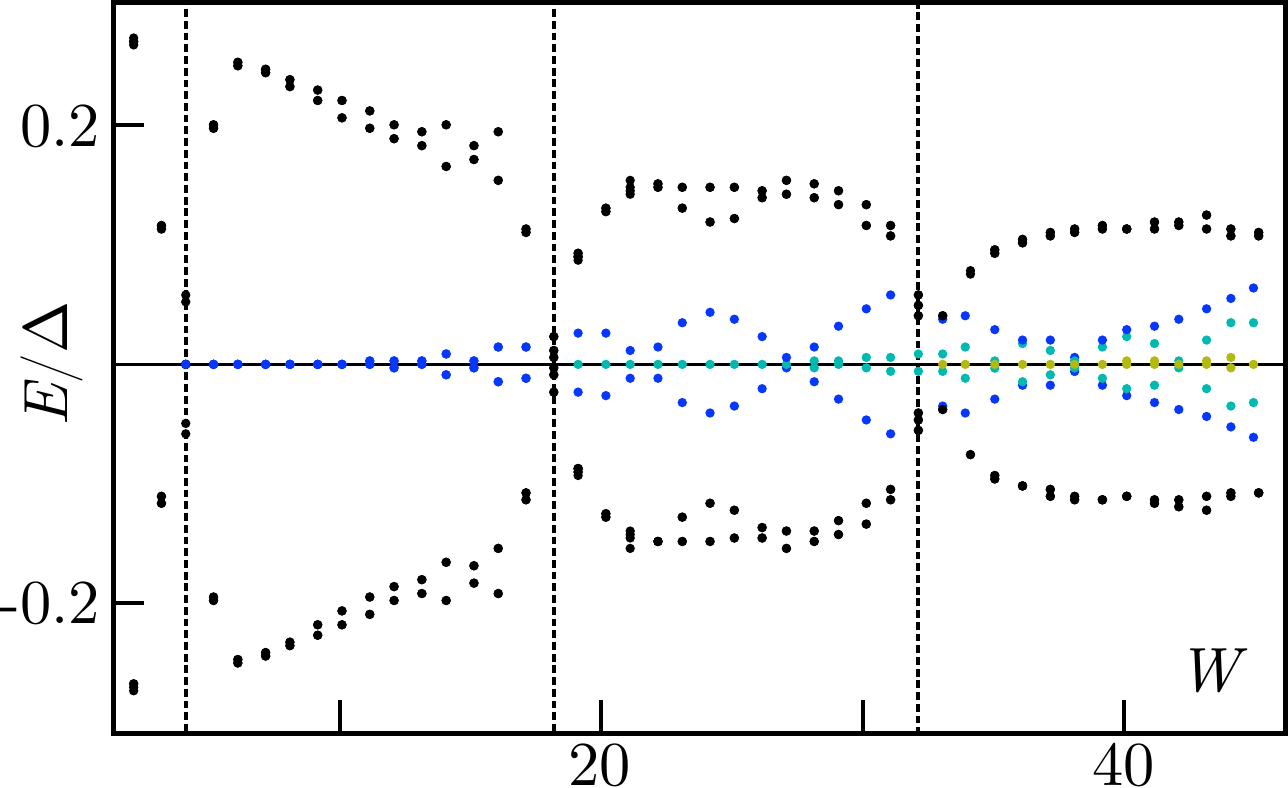}
	\caption{Energy of lowest states as function of $W$. Dashed lines mark the gap closing of the ABSB. After the first closing there are four zero-energy states, one Kramers pairs on each end of the normal region (blue). After the second gap closing there are two Kramers pairs of MBSs on each end (blue and turqoise). These two decoupled sets of MBSs come from the $n=1$ and $n=2$ subbands, as can be checked by counting the number of nodes of their wavefunction in $y$ direction. Since the smallest gap of the $n=1$ subband at the outer Fermi momemtum decreases with increasing $W$, the localization length increases and hence the zero-energy states are split away from $E=0$. This behavior repeats after the next gap closing where the $n=3$ band undergoes a topological phase transition and a third Kramers pair of MBS emerges (green). Parameters are: $N_x=800$, $N_y=350$, $\alpha_1/t=-\alpha_{\bar{1}}/t=0.07$, $\Delta=E_{so}/2$, and $\Gamma=3 \Delta$.}
	\label{Multiple_MBS}
\end{figure}

As has become clear from the discussion above, for fixed coupling strength $\Gamma$ and increasing width $W$, the lowest lying ABSB will change its `orbital content' around $k_x=0$ and will have contributions from transverse subbands with larger $n$ while the contributions from subbands with smaller $n$ are `pushed' to higher values of $k_x$ [see Fig. \ref{Spectra_transverse}]. Each time the band gap closes, a new transverse subband with higher $n$ undergoes a topological phase transition with the emergence of a new Kramers pair of MBSs [see Fig. \ref{Multiple_MBS}]. The wavefunctions of the MBSs show that they are localized on one end of the normal region as is expected. In the $y$-direction, however, the states that emerge for larger $W$ have $n$ peaks, revealing that they stem from the topological phase transition of the corresponding subband. 

Having discussed the physics around $k_x=0$, it is also important to look at finite momenta, as the localization length of the MBSs is determined by the smallest gap. Gaps at higher momenta, where the band has contributions from lower subbands, become smaller with increasing $W$ and thus, the localization length of MBSs coming from these subbands increases \cite{diego,prada}. This delocalization leads to a larger overlap of their wavefunction and thus they split in energy [see Fig. \ref{Multiple_MBS}]. Typically, when a new Kramers pair of MBSs emerges, the MBSs coming from subbands with lower $n$ are already split due their overlap, and there is only one Kramers pair of MBSs at each end that is at zero-energy. Thus, disorder induced coupling between an even number of Kramers pairs is suppressed since they are no longer energetically degenerate. As we have verified numerically, the phases with an even number of Kramers pairs of MBSs are indeed quite robust against potential disorder. Only in the limit where the junction is much longer in the $x$-direction than any of the localization lengths, the MBSs on opposite ends do not overlap, and one can always find a term that gaps out all MBSs without violating any symmetries. Strictly speaking only the phases with an odd number of Kramers pair of MBSs are topologically protected and thus stable against any symmetry-respecting perturbations.

{\it Breaking time-reversal symmetry.} 
 If TRS is broken by a Zeeman field along the $x$-direction, the system behaves similarly to what is known in topological double nanowires setups \cite{Schrade2017}. Increasing $\Delta_Z$ continuously from zero brings the system into a trivial phase until a critical field value is reached and the gap of the ABSBs closes. Increasing $\Delta_Z$ further leads to a reopening of the gap and the system enters a topological phase with a total of two MBSs, one MBS at each end of the normal region, see SM \cite{SM}. Importantly, there exist regions where the topological phase can be reached with relatively small Zeeman fields [see Fig. \ref{PhaseDiagZ}].

\begin{figure}
	\centering
	\includegraphics[width=.75\columnwidth]{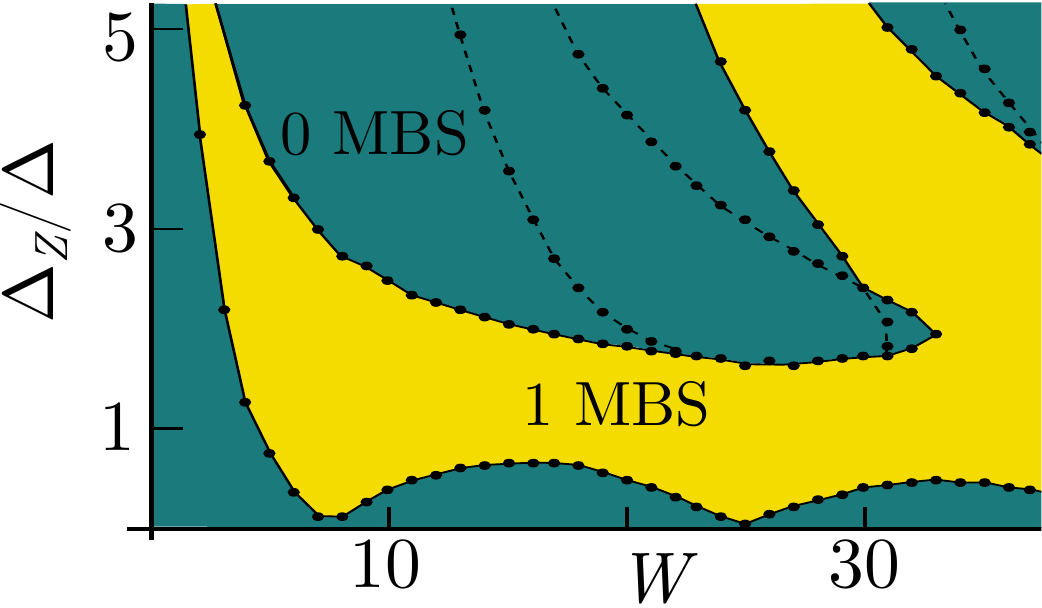}
	\caption{Phase diagram as function of in-plane Zeeman field $\Delta_Z$ and $W$. The Zeeman field is aligned along the $x$-direction. The solid black lines mark the topological phase transitions between the trivial (green) and the topological (yellow) phase, which harbors one MBS at each end of the normal region. Dashed lines mark closings of the gap of the ABSBs, which do not correspond to topological phase transitions. The phase boundaries were obtained for the same numerical parameters as in Fig. \ref{PhaseDiag}b and for $\Gamma=1.5\Delta$.}
	\label{PhaseDiagZ}
\end{figure}

{\it Acknowledgements.} We acknowledge helpful discussions with Olesia Dmytruk, Silas Hoffman, and Christopher Reeg. This work was supported by the Swiss National Science Foundation and NCCR QSIT. This project received funding from the European Union's Horizon 2020 research and innovation program (ERC Starting Grant, grant agreement No 757725).

\bibliographystyle{unsrt}

\onecolumngrid
\newpage
\vspace*{1cm}
\begin{center}
	\large{\bf Supplemental Material to `Time-Reversal Invariant Topological Superconductivity \\in Planar Josephson Bijunction' \\}
\end{center}
\begin{center}
	Yanick Volpez, Daniel Loss, and Jelena Klinovaja\\
	{\it Department of Physics, University of Basel, Klingelbergstrasse 82, CH-4056 Basel, Switzerland}
\end{center}
\vspace*{1cm}
\onecolumngrid
\setcounter{equation}{0}
\setcounter{figure}{0}

\section{Appendix A: Discretized lattice models}\label{TBM}

In the main text we present numerical results for the semi-infinite and the finite geometries in which the Rashba bilayer system is assumed to have a rectangular shape. In this section we explicitly discretize the total Hamiltonian $H$ defined by Eqs. (1)-(3) of the main text.

\subsection*{\it Semi-infinite geometry}

In the semi-infinite geometry, we assume, without loss of generality, that the system is translationally invariant along the $x$ and finite along the $y$ direction with the length $L_y=(N_y-1)a$, where $N_y$ is the number of lattice sites in $y$-direction and $a$ the lattice constant. The width of the left superconducting region is given by $L_1=(N_1-1)a$ and the width of the normal region is $W=(N_W+1)a$. The total Hamiltonian for the semi-infinite geometry is given by $\bar{H}'=\sum_{k_x}[\bar{H}'_1(k_x)+\bar{H}'_{\bar{1}}(k_x)+\bar{H}'_{\Gamma}(k_x)+\bar{H}'_D(k_x)]$ with
\begin{align}
\bar{H}'_{\tau}(k_x) &=\sum_{m} \Big\{ \sum_{\sigma} \Big(-t c^{\dagger}_{k_x \tau (m+1) \sigma} c_{k_x\tau m \sigma} + [-t \cos(k_x a) +\mu_{\tau}/2+2t] c^{\dagger}_{k_x \tau m \sigma} c_{k_x\tau m \sigma} + \text{H.c.} \Big) \nonumber \\
&\hspace{.5cm}+ \tau \tilde{\alpha}\Big[ i(c^{\dagger}_{k_x \tau (m+1) \uparrow}c_{k_x\tau m \downarrow} - c^{\dagger}_{k_x \tau (m-1) \uparrow} c_{k_x\tau m \downarrow})  + 2i \sin(k_x a) c^{\dagger}_{k_x \tau m \uparrow} c_{k_x\tau m \downarrow} + \text{H.c.} \Big] \Big\}, \nonumber \\
\bar{H}'_{\Gamma}(k_x) &= \Gamma \sum_{\sigma} \sum_{m=N_1+1}^{N_1+N_W} \Big( c^{\dagger}_{k_x 1 \sigma m}c_{k_x \bar{1} \sigma m} + \text{H.c.}\Big), \nonumber \\
\bar{H}'_{\Gamma'}(k_x) &= \Gamma' \sum_{\sigma} \left[ \sum_{m=1}^{N_1} \Big( c^{\dagger}_{k_x 1 \sigma m}c_{k_x \bar{1} \sigma m} + \text{H.c.}\Big)+ \sum_{m=N_1+N_W+1}^{N_y} \Big( c^{\dagger}_{k_x 1 \sigma m}c_{k_x \bar{1} \sigma m} + \text{H.c.}\Big) \right]\nonumber \\
\bar{H}'_D(k_x) &= \frac{1}{2} \sum_{\tau} \sum_{\sigma, \sigma'} \left[\sum_{m=1}^{N_1} \Big(\Delta e^{-i \phi/2} c^{\dagger}_{k_x \tau m \sigma} [i \sigma_2]_{\sigma \sigma'} c^{\dagger}_{-k_x \tau m \sigma'} + \text{H.c.}\Big) + \sum_{m=N_1+N_W+1}^{N_y} \Big(\Delta e^{i \phi/2} c^{\dagger}_{k_x \tau m \sigma} [i \sigma_2]_{\sigma \sigma'} c^{\dagger}_{-k_x \tau m \sigma'} + \text{H.c.}\Big)\right].
\end{align}
The operator $c^{\dagger}_{k_x \tau \sigma n}$ creates an electron with momentum $k_x$ and spin projection $\sigma$ (along $z$-axis) in the layer $\tau$ at the lattice site $m$. Here, $t$ is the amplitude for a hopping process between two neighboring lattice sites used to set the effective mass as $t=\hbar^2/(2m a^2)$.
The spin-flip hopping amplitude $\tilde{\alpha}$ is related to the SOI parameter by $\tilde{\alpha}=\alpha/2a$. The spin-orbit energy $E_{so}$ is given by $E_{so} = \tilde{\alpha}^2/t$ \cite{Diego, SOI}. We use this Hamiltonian to obtain the spectra and wavefunction in Fig. (2) in the main text, and to determine the position of the gap closing of the ABSBs.

\subsection*{\it Finite rectangular geometry}

In the finite geometry we assume the system to be finite in both $x$ and $y$ directions and of size $L_x \times L_y=(N_x-1)(N_y-1) a^2$. The indices $(n,m)$ label sites in the two-dimensional square lattice with $n \in \{1,\dots,N_x\}$ and $m \in \{1,\dots, N_y\}$. The total Hamiltonian $\bar{H}=\bar{H}_1+\bar{H}_{\bar{1}}+\bar{H}_{\Gamma}+\bar{H}_{\Gamma'}+\bar{H}_{D} $ in the finite geometry is then given by
\begin{align}\label{HamiltonianTB}
\bar{H}_{\tau} &= \sum_{n,m} \Big\{ \sum_{\sigma} \Big(-t_x c^{\dagger}_{\tau \sigma (n+1) m} c_{\tau \sigma n m} -t_y c^{\dagger}_{\tau \sigma n (m+1)} c_{\tau \sigma n m} + \frac{\mu_{\tau}+4t}{2} c^{\dagger}_{\tau \sigma n m} c_{\tau \sigma n m} + \text{H.c.} \Big) \nonumber \\ 
&\hspace{.5cm} + \tau \tilde{\alpha}\Big[ i(c^{\dagger}_{\tau \uparrow n (m+1) }c_{\tau \downarrow n m} - c^{\dagger}_{\tau \uparrow n (m-1)} c_{\tau \downarrow n m}) -(c^{\dagger}_{\tau \uparrow (n+1) m}c_{\tau \downarrow n m} - c^{\dagger}_{\tau \uparrow (n-1) m} c_{\tau \downarrow n m}) + \text{H.c.} \Big] \Big\}, \nonumber \\
\bar{H}_{\Gamma} &= \Gamma \sum_{n}\sum_{m=N_1+1}^{N_1+N_W} \Big( c^{\dagger}_{1 \sigma n m} c_{\bar{1} \sigma n m} + \text{H.c.}\Big), \nonumber \\
\bar{H}_{\Gamma'} &= \Gamma' \sum_{n}\left[\sum_{m=1}^{N_1} \Big( c^{\dagger}_{1 \sigma n m} c_{\bar{1} \sigma n m} + \text{H.c.}\Big)+\sum_{m=N_1+N_W+1}^{N_y} \Big( c^{\dagger}_{1 \sigma n m} c_{\bar{1} \sigma n m} + \text{H.c.}\Big) \right], \nonumber \\
\bar{H}_{\Delta} &= \frac{1}{2} \sum_{n} \sum_{\sigma, \sigma'} \left[ \sum_{m=1}^{N_1} \Big( \Delta e^{-i \phi/2} c^{\dagger}_{\tau \sigma n m} [i \sigma_2]_{\sigma \sigma'} c^{\dagger}_{\tau \sigma' n m} + \text{H.c.} \Big)+ \sum_{m=N_1+N_W+1}^{N_y} \Big( \Delta e^{i \phi/2} c^{\dagger}_{\tau \sigma n m} [i \sigma_2]_{\sigma \sigma'} c^{\dagger}_{\tau \sigma' n m} + \text{H.c.}\Big) \right].
\end{align}
The operator $c^{\dagger}_{ \tau \sigma n m}$ creates an electron with spin projection $\sigma$ in the layer $\tau$ at the lattice site $(n,m)$. Note that $\bar{H}'$ in the semi-infinite geometry can be obtained from $\bar{H}$ by applying the Fourier transformation in the $x$ direction. We use this Hamiltonian to obtain the spectra and MBS wavefunction in Figs. (4) and (5) in the main text, as well as to identify the different topological phases.
\begin{figure}
	\centering
	\includegraphics[width=.55\linewidth]{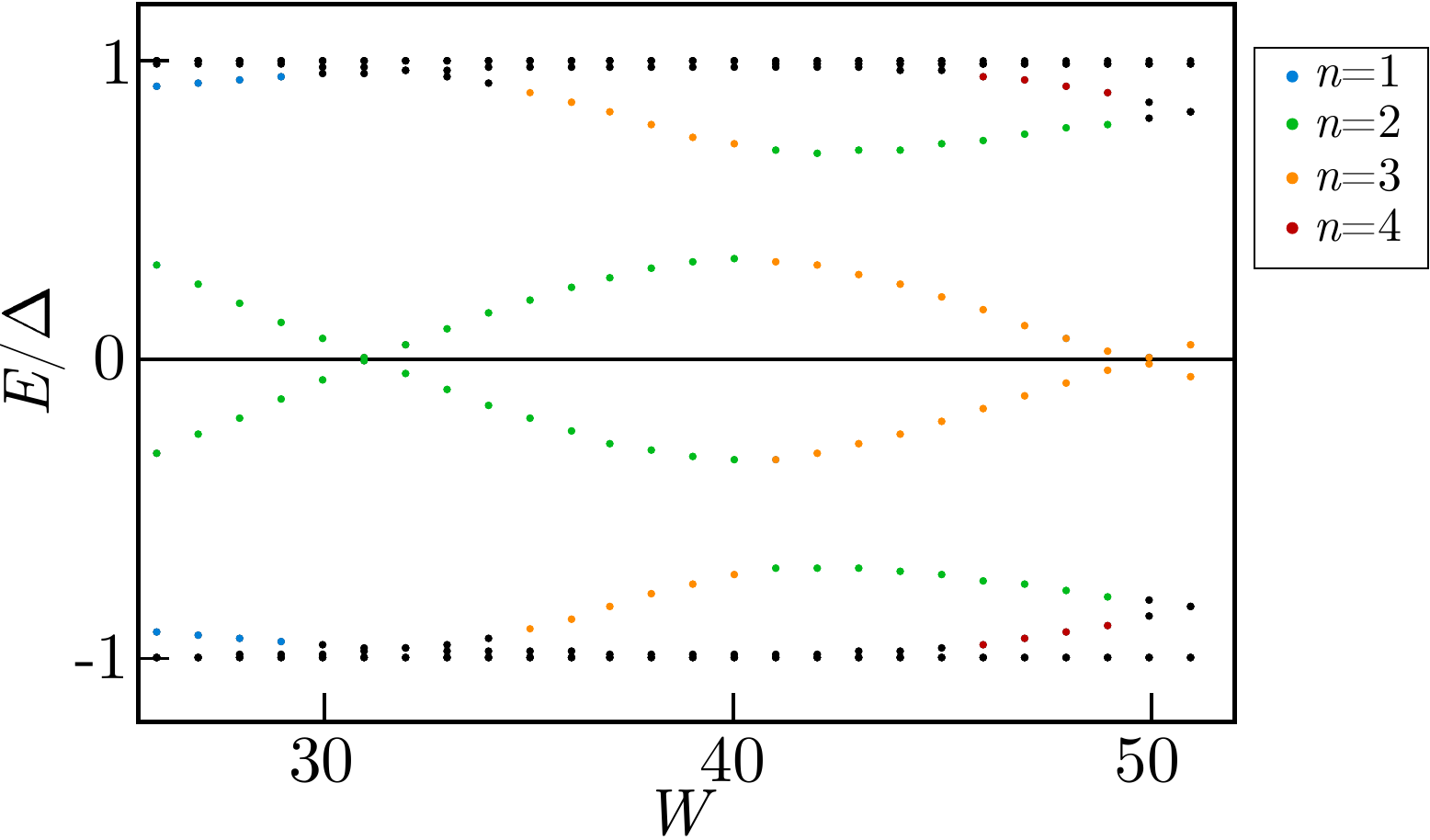}
	\caption{Spectrum in the semi-infinite geometry at $k_x=0$ as a function of $W$. The lowest ABSB with positive energy has a dominant contribution coming from the $n=2$ transverse subband (green dots). The gap closes at $W=30$, which corresponds to the topological phase transition of the $n=2$ subband. As $W$ is increased, the $n=3$ subband enters the superconducting gap and the second lowest ABSB in energy has its dominant contribution from this subband (orange dots). At $W=39$ the dominant contribution is switched and the lowest ABSB has then mainly a contribution from the $n=3$ subband. At $W=49$ the gap closes, which corresponds to the topological phase transition of the $n=3$ subband. Numerical parameters are: $N_x=1500$, $\alpha_1/t=-\alpha_{\bar{1}}/t=0.066$, $\Delta=E_{so}/2$, $\Gamma/\Delta=1.5$, and $V=t/10$.}
	\label{SpectrumKzero}
\end{figure}

\section*{Appendix B: Hybridization of transverse subbands}
\begin{figure}
	\centering
	\includegraphics[width=0.45\linewidth]{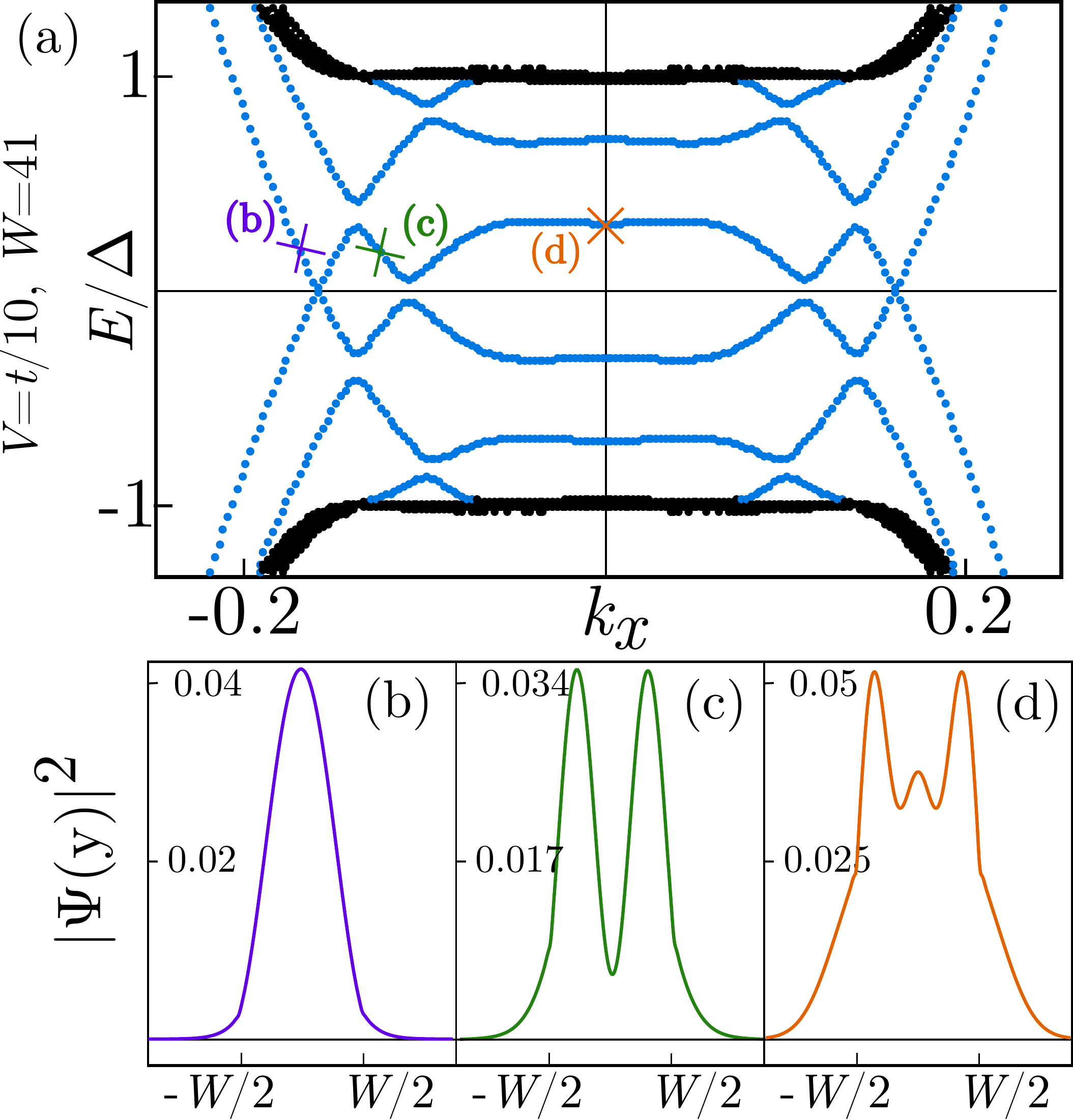}
	\caption{Spectrum in the semi-infinite geometry for $W=41$, $\Gamma/\Delta=1.5$ and $V=t/10$. At this point there are three ABSBs (blue dots) in the superconducting gap (black dots). The probability density of the wavefunction of the lowest ABSB at (b) $k_x=-0.17$ (purple), (c) $k_x=-0.12$ (green), and (d) $k_x=0$ (orange) shows that the band has dominant contributions from lower subbands [(b) $n=1$ and (c) $n=2$] at finite momenta. Numerical parameters are the same as in Fig. \ref{SpectrumKzero}.}
	\label{SpectrumFiniteK}
\end{figure}

As discussed in the main text the 'orbital content' of the ABSBs depends on the momentum $k_x$ and changes as a function of $W$. When $W$ is increased more transverse subbands of the normal region move into the superconducting gap and as a consequence more ABSBs are formed. Generally, the ABSBs in the tunnel coupled Josephson bijunction are gapped. In the semi-infinite geometry, the gap only closes at $k_x=0$ (the momentum along the junction) for certain combinations of $k_{so}W$ and $\Gamma/\Delta$. Examining the wavefunction of the ABS at a fixed momentum reveals which transverse subband is dominating at that point. In order to better understand the topological phase transitions of several transverse subbands, we consider the lowest ABSB at $k_x=0$, $\Gamma/\Delta=1.5$, and when $W$ is varied from $W=26$ to $W=51$ [see Fig. \ref{SpectrumKzero}]. For $W=26$ the dominant contribution is coming from the second transverse subband. As $W$ is increased, the gap closes at $W=31$ and then reopens as $W$ is increased further. This corresponds to the topological phase transition of the second transverse subband. Since the width of the normal region is increased, a new ABSB forms and the second lowest ABSB then has a dominant contribution from the third transverse subband. At some point the second and third transverse subband hybridize, such that the dominant contribution of the lowest ABSB is switched (at $W=40$) and the situation is now opposite. Increasing $W$ further, the gap of the lowest ABSB then closes at $W=50$, which corresponds to the topological phase transition of the third transverse subband.
While for the lowest ABSB the main contribution at $k_x=0$ is coming from higher transverse subbands as $W$ is increased, the contribution of the lower transverse subbands appears at finite momenta [see Fig. \ref{SpectrumFiniteK}]. The local minima at these finite momenta, where the band has dominant contributions from lower subbands, determines the localization lengths of potential bound states \cite{bound}. For $W=41$, the $n=1$ and the $n=2$ subbands are in the topological phase [see Fig. \ref{SpectrumKzero}], and as the gaps around $|k_x|=0.17$ ($n=1$) and around $|k_x|=0.12$ ($n=2$) are quite small [see Fig. \ref{SpectrumFiniteK}], the localization length of the MBS become quite large. We note that these localization lengths can be decreased if zigzag-shaped boundaries of the SNS interface are considered \cite{anton}.
This explains the behavior discussed in Fig. 5 in the main text.

\section*{Appendix C: ABS wavefunction and Topological Phase Transition Criterion}
In this section we give a detailed derivation of the ABS wavefunction present in the coupled Josephson junction setup. As explained in the main text, we assume the system to be translationally invariant along the $x$ direction (along the junction), and finite in the $y$ direction. We then focus on the subgap states at $k_x=0$ with energy $E<\Delta$ and the time-reversal invariant case $\phi=\pi$. The BdG equations for this problem then read
\begin{align}\label{BdG1}
\left[	-\eta_3\frac{\hbar^2}{2m} \partial_y^2 - i \tau \alpha \partial_y \sigma_1 + \bar{\Gamma}(y) \tau_1 \eta_3 -\Delta(y)\eta_2 \sigma_2 \right] \tilde{\Phi}_E(y)=E\tilde{\Phi}_E(y),
\end{align}
where $\bar{\Gamma}(y)=\Gamma \theta(W/2-|y|)$, and $\Delta(y)=\text{sgn}(y)\Delta\theta(|y|-W/2)$ and without loss of generality we assume $\Gamma>0$ and $\Delta>0$, as discussed in the main text. The wavefunction $\bar{\Phi}_E(y)$ is given by $\bar{\Phi}_E(y)=[u_{E,1 \uparrow}(y),u_{E,1 \downarrow}(y),v_{E,1 \uparrow}(y),v_{E,1 \downarrow}(y),u_{E,\bar{1} \uparrow}(y),u_{E,\bar{1} \downarrow}(y),v_{E,\bar{1} \uparrow}(y),v_{E,\bar{1} \downarrow}(y)]$. By inspecting Eq. \eqref{BdG1} one can see that there are actually two decoupled sets of equations, one only involving $\Phi^{(1)}_E(y)= [u_{E,1 \uparrow}(y),v_{E,1 \downarrow}(y),u_{E,\bar{1} \uparrow}(y),v_{E,\bar{1} \downarrow}(y)]$ and the other only involving $\Phi^{(2)}_E(y)= [u_{E,1 \downarrow}(y),v_{E,1 \uparrow}(y),u_{E,\bar{1} \downarrow}(y),v_{E,\bar{1} \uparrow}(y)]$. Since these two sectors are decoupled the spectrum is two-fold degenerate, and it is enough to focus on one sector in order to derive the zero-energy condition. Without loss of generality, we solve the BdG equations for $\Phi^{(1)}_E(y)$ and we drop from now on the superscript. First, we perform a unitary transformation $U=e^{i \pi \sigma_2/4}$, which maps $\sigma_1 \to \sigma_3$ and $\sigma_3 \to -\sigma_1$. With these preparations the reduced BdG equations read
\begin{align}
\left( \begin{array}{cccc}-\hbar^2/2m \partial_y^2 - i \alpha \partial_y & 0 & \Gamma(y) & \Delta(y) \\ 0 & \hbar^2/2m \partial_y^2 + i \alpha \partial_y & -\Delta(y) & -\Gamma(y) \\ \Gamma(y) & -\Delta^*(y) & -\hbar^2/2m \partial_y^2 + i \alpha \partial_y & 0 \\ \Delta^*(y) & -\Gamma(y) & 0 & \hbar^2/2m \partial_y^2 - i \alpha \partial_y \end{array} \right) \Phi(y)=0,
\end{align}
which are then solved in the normal region $|y|<W/2$, in the left superconducting region $y<-W/2$, and in the right superconducting region $y>W/2$. The general solution in the normal region is given by
\begin{align}
\Phi^{(N)}(y) &= a_1 \left[0, -\frac{2 E_{so} \lambda_-}{\Gamma} \left(\lambda_-+\sqrt{2}\right),0,1 \right] e^{-i \sqrt{2} \lambda_- k_{so}y} \nonumber \\
&+a_2 \left[0, -\frac{2 E_{so} \lambda_-}{\Gamma} \left(\lambda_- - \sqrt{2} \right),0,1 \right] e^{i \sqrt{2} \lambda_- k_{so}y} \nonumber \\
&+a_3 \left[0, -\frac{2 E_{so} \lambda_+}{\Gamma} \left(\lambda_+ + \sqrt{2} \right),0,1 \right] e^{-i \sqrt{2} \lambda_+ k_{so}y } \nonumber \\
&+ a_4 \left[0, -\frac{2 E_{so} \lambda_+}{\Gamma} \left(\lambda_+ -\sqrt{2}\right),0,1 \right] e^{i \sqrt{2} \lambda_+ k_{so}y } \nonumber \\
&+a_5 \left[ -\frac{2 E_{so} \lambda_-}{\Gamma} \left(\lambda_- + \sqrt{2}\right),0,1,0 \right] e^{-i \sqrt{2} \lambda_- k_{so}y} \nonumber \\
&+a_6 \left[ -\frac{2 E_{so} \lambda_-}{\Gamma} \left(\lambda_- - \sqrt{2}\right),0,1,0 \right] e^{i \sqrt{2} \lambda_- k_{so}y} \nonumber \\
&+a_7 \left[-\frac{2 E_{so} \lambda_+}{\Gamma} \left(\lambda_+ + \sqrt{2}\right),0,1,0 \right] e^{-i \sqrt{2} \lambda_+ k_{so}y} \nonumber \\
&+ a_8 \left[ -\frac{2 E_{so} \lambda_+}{\Gamma} \left(\lambda_+ - \sqrt{2}\right),0,1,0 \right] e^{i \sqrt{2} \lambda_+ k_{so}y}, \nonumber \\
\end{align}
where $\lambda_{\pm} = \sqrt{1 \pm \sqrt{1+\frac{\Gamma^2}{4 E_{so}^2}}}$. Since we are interested in subgap solutions with $E<\Delta$, the states have to be decaying in the superconducting regions. For the left superconducting region $y<-W/2$ the solution has to decay for $y\to - \infty$. We find,
\begin{align}
\Phi^{(L)}(y)&=a_9 \left[1,1,0,0 \right]e^{-i k_{so} y \left(1+\sqrt{1+i\frac{\Delta}{E_{so}}} \right)} + a_{10} [-1,1,0,0] e^{i k_{so}y \left(-1 + \sqrt{1 - i \frac{\Delta}{E_{so}}} \right)} \nonumber \\
&+a_{11} [0,0,-1,1] e^{i k_{so}y \left(1 + \sqrt{1-i \frac{\Delta}{E_{so}}} \right)} +a_{12} [0,0,1,1] e^{-i k_{so}y \left(-1 + \sqrt{1+i \frac{\Delta}{E_{so}}}\right)}.
\end{align}
Analogously one finds the wavefunction in the right superconducting region, which decays for $y \to \infty$,
\begin{align}
\Phi^{(R)}(y) &= a_{13} [0,0,-1,1] e^{i k_{so} y \left(1 + \sqrt{1+ i \frac{\Delta}{E_{so}}}\right)} + a_{14} [0,0,1,1] e^{-i k_{so}y \left(-1 +\sqrt{1	-i \frac{\Delta}{E_{so}}}\right)} \nonumber \\
&+a_{15}[1,1,0,0] e^{-i k_{so}y \left(1 + \sqrt{1 - i \frac{\Delta}{E_{so}}}\right)} + a_{16}[-1,1,0,0] e^{i k_{so} y \left(-1 + \sqrt{1 + i \frac{\Delta}{E_{so}}}\right)}.
\end{align}
The total ABS wavefunction then reads $\Phi(y) = \Phi^{(L)}(y) \theta(-y-W/2) + \Phi^{(N)}(y) \theta(W/2-|y|) + \Phi^{(R)}(y) \theta(y-W/2)$. One has to impose the boundary conditions at the two interfaces,
\begin{align}
\Phi^{(R)}(-W/2) - \Phi^{(N)}(-W/2) \overset{!}{=}0, \\
\partial_y \Phi^{(R)}(-W/2) - \partial_y \Phi^{(N)}(-W/2) \overset{!}{=}0, \\
\Phi^{(L)}(W/2) - \Phi^{(N)}(W/2) \overset{!}{=}0, \\
\partial_y \Phi^{(L)}(W/2) - \partial_y\Phi^{(N)}(W/2) \overset{!}{=}0,
\end{align}
which can be recast in matrix form as
\begin{equation}\label{Determinant1}
M a =0,
\end{equation}
where $a=(a_1, \dots,a_{16})$ is the vector of linear coefficients. Non-trivial solutions exist only if the matrix $M$ is singular,{\it i.e.} $\text{det}(M)=0$. The solution of this equation then yields the condition on the system parameters ($k_{so}W$, $\Gamma$, $\Delta$) under which a zero-energy ABS exists. This is the topological phase transition criterion we are looking for. As can be seen above, the presence of the square root factors makes it impossible to find an analytical closed form expression. However, in the strong SOI limit, {\it i.e.} $\Gamma$, $\Delta \ll E_{so}$, these square roots can be expanded and thus the matrix $M$ and the equation Eq. \eqref{Determinant1} are expanded in the strong SOI limit, and can be brought into form
\begin{equation}\label{Determinant2}
\text{det}(M) = f^{(0)}(k_{so}W,\Gamma,\Delta) +\frac{f^{(1)}(k_{so}W,\Gamma,\Delta)}{E_{so}} + \frac{f^{(2)}(k_{so}W,\Gamma,\Delta)}{E_{so}^2} \overset{!}{=}0
\end{equation}
In doing so, we find
\begin{align}
f^{(0)}(k_{so}W,\Gamma,\Delta) &= 0 \nonumber \\
f^{(1)}(k_{so}W,\Gamma,\Delta) &= 0 \nonumber \\
f^{(2)}(k_{so}W,\Gamma,\Delta) &= -\frac{\cos(k_{so}W)\left[\Delta^2- k_{so}W \Gamma^2 \tan(k_{so}W) \right]}{\Gamma^2}
\end{align}
Thus for $\Gamma>0$, Eq. \eqref{Determinant2} can only be fulfilled if
\begin{equation}
\frac{\Gamma^2}{\Delta^2} = \frac{\cot(k_{so}W)}{k_{so}W}
\end{equation}
One can see that the system only has zero-energy bound states when the cotangent is positive, since throughout the derivation it was assumed that $\Gamma>0$.

\section*{Appendix C: Effect of Breaking Time-Reversal Symmetry}

In the main text we presented the results for the TRI topological superconducting phase and briefly mentioned the effect of breaking time-reversal symmetry (TRS). In this section we give a more detailed discussion of the TRS broken phase.

In our system, TRS can be broken by either deviating from a $\pi$-junction, {\it i.e} $\phi=\pi+\delta \phi$, or by applying a Zeeman field. The TRS breaking term coming from the deviation from a $\pi$-junction, acts in the same way on all states independently of their spin-structure. This results in trivially gapping out the MBS, and the system is always in a trivial phase.

We take into account the effect of applying a Zeeman field along the junction (the $x$-direction), by adding the term
\begin{equation}
H_Z = \sum_{\tau,\sigma,\sigma'} \int d {\bf r} \ \bar{\Delta}_Z(y) \Big(\psi_{\tau \sigma}({\bf r})^{\dagger} [\sigma_1]_{\sigma \sigma'} \psi_{\tau \sigma'}({\bf r}) + \text{H.c.} \Big),
\end{equation}
where $\bar{\Delta}_Z(y)=\Delta_Z \theta(W/2-|y|)$, and we assume without loss of generality that $\Delta_Z \geq0$. We find that the Zeeman field leads to a closing of the gap in the ABSBs at $k_x=0$ for some critical fields. Starting with $\Delta_Z=0$ for some fixed width $W$, one starts in the TRI topological phases discussed in the main text, which then become trivial under the application of the Zeeman field, since the Kramers pairs of MBS are gapped out. Increasing $\Delta_Z$ furhter until the ABSB gap closes and reopens the first time, the system enters a topological phase with one MBS at each end of the junction [see Fig. \ref{PhaseDiagZ}]. After the second gap closing and reopening the system has an even number of MBS, which are not protected against disorder, and indeed we observe that they split as a function of disorder strength, and hence, this region is also trivial.

\begin{figure}
	\centering
	\includegraphics[width=.35\linewidth]{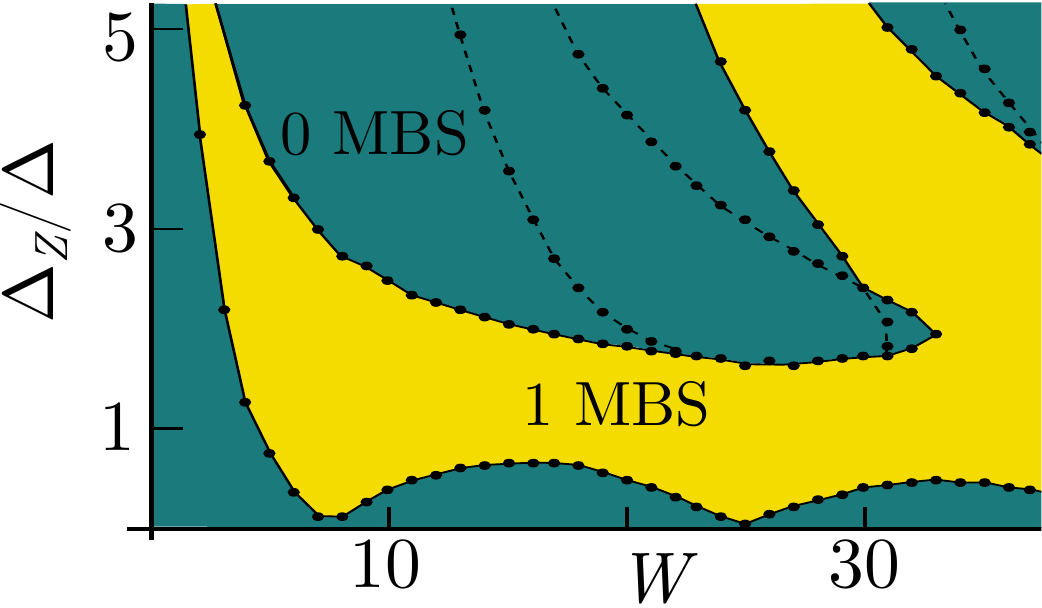}
	\caption{Phase diagram as a function of $\Delta_Z/\Delta$ and the width $W$ for $\Gamma=1.5\Delta$. The black dots mark the points where the gap in the ABSBs is closed. The yellow (green) regions mark the topological (trivial) phases. The topological phase is characterized by the existence of one MBS at each end of the junction. Numerical parameters are $N_y=10000$, $\alpha_1/t=-\alpha_{\bar{1}}/t=0.07$, $\Delta/t=0.01$, $\Gamma'=0$, and $\Gamma=1.5\Delta$.}
	\label{PhaseDiagZ}
\end{figure}
\newpage
\bibliographystyle{unsrt}

\begin{thebibliography}{}
\bibitem{Footnote1}
We note that crossed Andreev superconducting pairing \cite{car1,car2,car3,car4,car5,car6} is suppressed in this setup since the signs of the SOI couplings $\alpha_{\tau}=|\boldsymbol{\alpha}_{\tau}|$ are opposite.
\end{thebibliography}

 \end{document}